\documentclass[11pt]{article}
\usepackage{graphicx} 
\usepackage{amssymb,amsthm}  
\usepackage{cite}   

\begin{document}

\def \d {{\rm d}}
\def \U {{\cal U}}
\def \V {{\cal V}}
\def \H {{\cal H}}
\def \M {{\cal M}}
\def \e {\varepsilon}
\def \E {{\bf e}}
\def \m  {{\bf m}}
\def \k  {{\bf k}}
\def \uu {{\bf u}}
\def \R  {{\cal R}}
\def \I  {{\cal I}}
\def \A  {{\cal A}}
\def \C  {{\cal C}}
\newcommand{\be}{\begin{equation}}
\newcommand{\ee}{\end{equation}}
\newcommand{\beqn}{\begin{eqnarray}}
\newcommand{\eeqn}{\end{eqnarray}}
\newcommand{\pa}{\partial}
\newcommand{\ba}{\begin{array}}
\newcommand{\ea}{\end{array}}

\newcommand{\pp}{{\it pp\,}-}


\def \E {{\bf e}}
\def \m  {\mbox{\boldmath{$m$}}}
\def \mc  {{\bf\bar m}}
\def \bl  {\mbox{\boldmath{$l$}}}
\def \k  {{\bf k}}
\def \uu {{\bf u}}
\def \bn {\mbox{\boldmath{$n$}}}

\def \bt  {\mbox{\boldmath{$t$}}}
\def \bs  {\mbox{\boldmath{$s$}}}

\def \ep  {\mbox{\boldmath{$\epsilon$}}}
\def \del  {\mbox{\boldmath{$\delta$}}}

\def \a {\alpha}
\def \ac {\bar \alpha}
\def \b {\beta}
\def \bc {\bar \beta}
\def \g {\gamma}
\def \gc {\bar \gamma}
\def \e {\varepsilon}
\def \ec {\bar \varepsilon}
\def \kk {\kappa}
\def \kkc {\bar \kappa}
\def \l {\lambda}
\def \lc {\bar \lambda}
\def \mm {\mu}
\def \mmc {\bar \mu}
\def \n {\nu}
\def \nc {\bar \nu}
\def \p {\pi}
\def \pc {\bar \pi}
\def \rr {\rho}
\def \rc {\bar \rho}
\def \s {\sigma}
\def \sc {\bar \sigma}
\def \t {\tau}
\def \tc  {\bar \tau}

\def \dec {\bar \delta}
\def \de {\delta}
\def \D {\Delta}

\def \phoo {\Phi_{11}}
\def \pho {\Phi_1}
\def \phoc {\bar\Phi_1}
\def \phtt {\Phi_{22}}
\def \pht {\Phi_2}
\def \phtc {\bar\Phi_2}
\def \pn {\Psi_0}
\def \po {\Psi_1}
\def \pt {\Psi_2}
\def \ptt {\Psi_3}
\def \pf {\Psi_4}

\newtheorem*{theorem}{Theorem}

\title{Bel-Debever criteria for the classification of the Weyl tensors in higher dimensions}

\addvspace{1cm}

\author{
Marcello Ortaggio\thanks{E--mail: {\tt ortaggio(at)math(dot)cas(dot)cz}}
\\ Institute of Mathematics, Academy of Sciences of the Czech Republic,\\
Zitn\' a 25, 115 67 Prague 1, Czech Republic
 \\ \\
}

\date{\today}

\maketitle
\begin{abstract}
An extension to higher dimensions of the Bel-Debever characterization of the Weyl tensor is considered. This provides algebraic conditions that uniquely determine the multiplicity of a Weyl aligned null direction (WAND), and thus the principal Weyl type, in a frame independent way. The specification of several ``subtypes'' is also encompassed by the criteria. We further comment on a Cartan-like geometrical interpretation of WANDs in terms of their invariance properties under parallel transport around infinitesimal loops. As a result, restrictions on the algebraic types permitted in spacetimes that admit a recurrent/covariantly constant vector field are outlined.


\end{abstract}


\section{Introduction}

The Petrov classification of the Weyl tensor provides an invariant characterization of four-dimensional geometries and it thus fundamental to the study of exact solutions of Einstein's equations. It admits various formulations
that, while describing properties of different geometric objects (bivectors, null directions, spinors), are in fact equivalent (see, e.g., \cite{Stephanibook,penrosebook2,Hallbook} for reviews and references). 

In recent years, gravity in higher dimensions has become an active area of ongoing studies. In particular, an algebraic classification of the Weyl tensor in any $n>4$ spacetime dimensions has been presented in \cite{Coleyetal04,Milsonetal05}. It characterizes the Weyl tensor in terms of the possible existence of Weyl aligned null directions (WANDs) and their order of alignment. In four dimensions, this is equivalent to the standard Petrov classification formulated in terms of the multiplicity of principal null directions (PNDs) \cite{Milsonetal05}. Since, however, there exists several approaches to the Petrov classification when $n=4$, it is natural to investigate whether such alternative methods also extend to higher dimensions, and whether they are still equivalent to the scheme proposed in \cite{Coleyetal04,Milsonetal05}. Looking at the Weyl classification from a different viewpoint may help elucidate further aspects of its geometric significance. Possibly, it can also lead to alternative definitions of algebraically special types that may be convenient for specific calculations. 

In four dimensions, one definition of the Petrov types is formulated in terms of algebraic relations (contractions and wedge products) involving the Weyl tensor and a null vector field $\bl$ (for instance, $\bl$ is a PND iff $l_{[e}C_{a]bc[d}l_{f]}l^bl^c=0$). It was presented (with minor formal differences) in several classic papers, apparently first by Bel and Debever \cite{Bel59,Bel62,Debever59,Debever59b} (see also, e.g., \cite{Sachs61} and \cite{Stephanibook,Hallbook}) and it will be briefly reviewed in the following. 
An advantage of this method is that it does not require a null frame to be introduced, so that definitions of algebraic special types are manifestly frame-independent. 
This proved useful, for instance, in the analysis of the peeling properties of radiative spacetimes \cite{Sachs61}. Further, equations of the Bel-Debever type arise naturally in the study of spacetimes that admit a geometrically privileged (e.g., recurrent or covariantly constant) null vector field (see~\cite{Stephanibook,Hallbook} and the discussion to follow), thus providing an immediate constraint on the algebraic type of the corresponding Riemann/Weyl  tensor. 
 
It was observed in \cite{Milsonetal05} that the aforementioned four-dimensional PND condition carries over with no changes into higher dimensions, and can then be used as a definition of WANDs for any $n\ge 4$.\footnote{An important difference, though, is that in higher dimensions real WANDs need not exist, in general \cite{Coleyetal04,Milsonetal05}, as opposed to the well-known situation in four dimensions.} Indeed, the explicit calculation of the Weyl type of certain black rings \cite{PraPra05,OrtPra06} relied on that. 
However, in \cite{PraPra05} it was pointed out that the standard four-dimensional criteria \cite{Stephanibook,Hallbook} for algebraically special types are only necessary conditions in higher dimensions and do not thus provide a complete algebraic classification.\footnote{For example, type N $\Rightarrow$ $C_{abcd}l^d=0$, but $C_{abcd}l^d=0$ $\not\Rightarrow$ type N for $n>4$ \cite{PraPra05} (explicit simple examples are mentioned in \cite{OrtPraPra09}), whereas $n=4$ spacetimes are of type N (or O) $\Leftrightarrow$ $C_{abcd}l^d=0$ \cite{Stephanibook,Hallbook} -- cf.~also the following discussion.} It is the main purpose of this paper to present an extension of the Bel-Debever criteria that fully characterize the principal algebraic types of the Weyl tensor in arbitrary dimension $n\ge 4$ (section~\ref{sec_BelDebever}). Most ``subtypes'' (see p.~L39 of \cite{Coleyetal04} and the summary below) will be also described, and the well-known four dimensional results will be recovered for $n=4$. 
In addition, building on an early observation by Cartan \cite{Cartan22}, we provide a geometrical characterization of WANDs as null direction that are invariant under parallel transport around appropriate infinitesimal loops (section~\ref{sec_cartan}). Several Weyl types will thus be singled out by requiring the existence of null directions with certain invariance properties. Possible applications in the study of holonomy are then mentioned. Comments on the Weyl types compatible with invariance of non-null vectors will conclude the paper.

\section{Bel-Debever criteria}

\label{sec_BelDebever}

\subsection{Classification of the Weyl tensor in higher dimensions}

The classification of the Weyl tensor in higher dimensions was presented in \cite{Coleyetal04,Milsonetal05} and has since been discussed in several papers, see, e.g., \cite{PraPra05} and the review \cite{Coley08}. It suffices here to  summarize only the basic definitions needed in the following. One first sets up a frame consisting of two null vectors $\m_{(0)}=\bl$, $\m_{(1)}=\bn$ and $n-2$ orthonormal spacelike vectors $\m_{(i)}$ (from now on, $0,1$ and $i, j, \dots=2,\ldots,n-1$ are frame indices), with the only non-zero components of the flat frame metric defined by $\eta_{01}=1$ and $\eta_{ij}=\delta_{ij}$ (so that $\m^{(i)}=\m_{(i)}$), and the spacetime metric by 
$g_{a b}=2l_{(a}n_{b)} + \delta_{ij} m^{(i)}_a m^{(j)}_b$.
Then one considers the frame components of the Weyl tensor. 
If there exists a vector field $\bl$ such that $C_{0i0j}=0$, then $\bl$ is a WAND and the Weyl tensor is algebraically special. The possible further vanishing of components of lower boost order \cite{Coleyetal04,Milsonetal05} determines the order of alignment of $\bl$, and the (principal) Weyl type of a spacetime is 
defined according to the following scheme:\footnote{The secondary classification \cite{Coleyetal04} (which involves the alignment of a possible second WAND $\bn$) need not be discussed explicitly here since the Bel-Debever criteria can be applied to {\em any} null direction in order to determine its alignment.}
\be
\ba{ll}
 \mbox{Type I:} & \quad C_{0i0j}=0 \\ 
 \mbox{Type II:} & \quad C_{0i0j}=C_{010i}=C_{0ijk}=0  \\ 
 \mbox{Type III:} & \quad C_{0i0j}=C_{010i}=C_{0ijk}=C_{01ij}=C_{0101}=C_{0i1j}=C_{ijkl}=0  \\ 
 \mbox{Type N:} & \quad C_{0i0j}=C_{010i}=C_{0ijk}=C_{01ij}=C_{0101}=C_{0i1j}=C_{ijkl}=C_{101i}=C_{1ijk}=0. 
 \ea
 \label{types}
\ee
In the case of types II, III and N the corresponding WAND $\bl$ is called ``multiple''. 
The spacetime is conformally flat (type O) if the Weyl tensor vanishes identically. Note that in the above definitions there is some redundancy, for the Weyl tensor is tracefree (and has well-known index symmetries), so that \cite{Coleyetal04}
\beqn
 & & C_{0i0i}=0=C_{1i1i} , \qquad C_{010i}=C_{0jij}, \qquad C_{101i}=C_{1jij}, \nonumber \\
 & & 2C_{0(i|1|j)}=-C_{ikjk}, \qquad 2C_{0[i|1|j]}=C_{01ij}, \qquad 2C_{0101}=-C_{ijij}=2C_{0i1i} .
 \label{tracefree}
\eeqn
However, keeping the full set~(\ref{types}) without the simplifications coming from (\ref{tracefree}) enables one to analogously classify also Riemann-like tensors \cite{Milsonetal05}.

One can also define specific ``subtypes'' when additional conditions are met, namely \cite{Coleyetal04} (for each subtype it is understood that the previous conditions~(\ref{types}) for the corresponding principal type must also hold): Type I$_a$: $C_{010i}=0$; Type II$_a$: $C_{0101}=0$; Type II$_b$: $C_{0(i|1|j)}=C_{0101}\delta_{ij}/(n-2)$; Type II$_c$: the ``Weyl part'' of $C_{ijkl}$ vanishes; Type II$_d$: $C_{01ij}=0$; Type III$_a$: $C_{011i}=0$.
For type II some subtypes can also hold simultaneously, in which case we shall use the notation II$_{ab}$, II$_{abd}$, etc..

\subsection{Criteria for multiple WANDs}

As already mentioned, ref.~\cite{Milsonetal05} has proved that in any higher dimensions
\be
 \bl \mbox{ is a WAND } \Leftrightarrow \ l_{[e}C_{a]bc[d}l_{f]}l^bl^c=0 , 
\ee
exactly as in four dimensions. We have extended this to similar conditions characterizing all principal types~(\ref{types}) and most subtypes, as summarized in table~\ref{tab_BelDebever}. These were arrived at by considering the frame decomposition of the Weyl tensor \cite{Coleyetal04} and looking for suitable contractions/``wedge products'' with $\bl$ so as to single out components with specific boost weights, and requiring these to vanish. The resulting conditions can also be straightforwardly checked {\em a posteriori} by contraction with the vectors of an {\em arbitrary} frame adapted to $\bl$, as we now exemplify in the case of type II and N (the remaining cases can be dealt with similarly).

Type II is specified by the equation $l_{[e}C_{a]b[cd}l_{f]}l^b=0$, i.e., (recall $\bl=\m_{(0)}$) $l_{[e}C_{a]0[cd}l_{f]}=0$. Components of the l.h.s. of this equations are clearly non-zero only when $c,d,f$ take different values. In addition, any non-zero frame-projection must contain $\m^f_{(1)}=\bn^f$ (up to permutation of $f$ with $c$ or $d$). We are thus left with the 
two equations $l_{[e}C_{a]00i}=0$ and $l_{[e}C_{a]0ij}=0$. 
All frame projections of these are identically satisfied except the product with $\m^e_{(1)}\m^a_{(k)}$,
thus giving, respectively, $C_{k00i}=0$ and $C_{k0ij}=0$, Q.E.D. (The argument works also in the reversed direction, i.e. the vanishing of these frame components is also a sufficient condition for having $l_{[e}C_{a]b[cd}l_{f]}l^b=0$.)

For type N we have $C_{ab[cd}l_{e]}=0$. Similarly as above, this reduces to the two equations $C_{ab0i}=0$ and $C_{abij}=0$. By all possible contractions with the frame vectors, the first of these gives $C_{0i0j}=C_{010i}=C_{0i1j}=C_{0ijk}=0$. From the second equation we additionally obtain $C_{01ij}=C_{ijkl}=C_{1kil}=0$, which completes the proof.

\begin{table}[t]
\begin{center}
	\begin{tabular}{|l|l|c|}
		\hline
			 &&\\[-3mm]
			 Bel-Debever conditions & \ Vanishing Weyl components \  & \ \ Type \ \  \\
			 &&\\[-3mm]
			 \hline\hline &&\\[-10pt]
  $l_{[e}C_{a]bc[d}l_{f]}l^bl^c=0$ \hfill {\scriptsize (*)} & $C_{0i0i}$ & I$_{\ }$ \\[1pt]
   	\hline &&\\[-10pt]
	$l_{[e}C_{a]bcd}l^bl^c=0$ \hfill {\scriptsize (*)} & $C_{0i0j}$;  $C_{010i}$ &  I$_{a}$ \\[1pt]
		\hline\hline &&\\[-10pt]
	$l_{[e}C_{a]b[cd}l_{f]}l^b=0$ & $C_{0i0j}$; $C_{0ijk}$ ($C_{010i}$) & II$_{\ \ \ }$ \\[1pt]
		\hline &&\\[-10pt]
	$l_{[e}C_{a]b[cd}l_{f]}l^b=0$, & & \\ $C_{abcd}l^bl^c=0$ & $C_{0i0j}$; $C_{010i}$, $C_{0ijk}$; $C_{0101}$ &  II$_{a\ \ }$ \\[1pt]
		\hline
	$C_{ab[cd}l_{e]}l^b=0$ & $C_{0i0j}$; $C_{010i}$, $C_{0ijk}$; $C_{01ij}$ &  II$_{d\ \ }$ \\[1pt]
		\hline
		$l_{[e}C_{ab][cd}l_{f]}=0$ & $C_{0i0j}$; $C_{0ijk}$ ($C_{010i}$); $C_{ijkl}$ ($C_{0101}, C_{0(i|1|j)}$) &  II$_{abc}$ \\[1pt]
		\hline
	$C_{abc[d}l_{e]}l^c=0$ \hfill {\scriptsize (*)} & $C_{0i0j}$; $C_{010i}$, $C_{0ijk}$; $C_{0i1j}$ ($C_{0101}, C_{01ij}$) &  II$_{abd}$ \\[1pt]
	\hline
	$C_{abcd}l^d=0$ \hfill {\scriptsize (*)} & $C_{0i0j}$; $C_{010i}$, $C_{0ijk}$; $C_{01ij}$, $C_{0101}$, $C_{0i1j}$; $C_{101i}$ &  II$'_{abd}$ \\[1pt]
		\hline\hline
	$l_{[e}C_{ab][cd}l_{f]}=0$, & & \\ $C_{abc[d}l_{e]}l^c=0$ & $C_{0i0j}$; $C_{010i}$, $C_{0ijk}$; $C_{0i1j}$ ($C_{0101}, C_{01ij}$), $C_{ijkl}$ &  III$_{\ }$ \\[1pt] \hline
	$l_{[e}C_{ab][cd}l_{f]}=0$, & & \\ $C_{abcd}l^d=0$ & $C_{0i0j}$; $C_{010i}$, $C_{0ijk}$; $C_{01ij}$, $C_{0101}$, $C_{0i1j}$, $C_{ijkl}$; $C_{101i}$ &  III$_{a}$ \\[1pt] \hline\hline
$C_{ab[cd}l_{e]}=0$ & $C_{0i0j}$; $C_{010i}$, $C_{0ijk}$; $C_{01ij}$, $C_{0i1j}$ ($C_{0101}$), $C_{ijkl}$; $C_{1ijk}$ ($C_{101i}$) &  N \\[1pt] 	
		\hline    
		\end{tabular}
		\end{center}
		\caption{\small~Bel-Debever criteria in $n\ge 4$ dimensions in order of increasing specialization of the algebraic type of the Weyl tensor. For each row, the condition in the first two columns are equivalent (in the second column, the Weyl components are sorted by boost weight, components with different weight are separated by semicolon, and $\bl$ is used as a frame vector). They imply that the algebraic type is at least as special as indicated in the third column (it can be more special if additional conditions are  satisfied by $\bl$ or, in the case of the first two rows, if there exists another WAND, different from $\bl$, with higher order or alignement). Viceversa, if the type is the one given in the third column, the corresponding WAND must satisfy the conditions in the first two columns. Components that vanish only thanks to the tracefree property~(\ref{tracefree}) of the Weyl tensor (e.g., $C_{0ijk}=0\Rightarrow C_{010i}=0$) are given in round brackets: bearing this in mind, the Bel-Debever criteria can thus be used also to (partly) classify Riemann-like tensors.
Conditions marked by  {\scriptsize (*)} are those usually considered in four dimensions \cite{Stephanibook,Hallbook}, in which case the possible algebraic (Petrov) types are fewer, and one has indeed the following equivalences: I$_a\equiv$II, II$\equiv$II$_b\equiv$II$_c$, II$_{abd}\equiv$III, II$'_{abd}\equiv$III$_a\equiv$N (and II$_{abc}\equiv$II$_a$, but such a notation for this subtype is generally not used in four-dimensions). Note that II$\equiv$II$_c$, II$'_{abd}\equiv$III$_a$ and II$_{abd}\equiv$III hold also in five dimensions since $C_{ijkl}$ can be expressed in terms of $C_{0(i|1|j)}$ \cite{PraPraOrt07}.}
		\label{tab_BelDebever}
\end{table}

\subsection{Comments on the criteria}

The Bel-Debever criteria can be employed to obtain the order of alignment of any null direction. In a given spacetime, finding a WAND whose order of alignment is as large as possible determines the principal Weyl type. This is understood when giving the type in the third column of table~\ref{tab_BelDebever}. 

Discussing the secondary classification \cite{Coleyetal04} is straightforward and need not be done explicitly here, since one has only to study similar conditions for a possible second WAND $\bn$ (not parallel to $\bl$). For example, for type D one needs two multiple WANDs to exist, i.e.
\be
 \mbox{Type D:} \qquad l_{[e}C_{a]b[cd}l_{f]}l^b=0, \quad n_{[e}C_{a]b[cd}n_{f]}n^b=0 .
\ee

We do not claim that the presented criteria are the only possible characterization of the algebraic types, and indeed it is not difficult to find different but equivalent ones, at least in some cases (this is true also in four dimensions \cite{Stephanibook,Hallbook}, where one can alternatively also formulate them in terms of the self-dual Weyl tensor). For instance, type I$_a$ can equivalently be defined by $C_{abcd}l^bl^d=\lambda l_al_c$.

In addition to those of \cite{Coleyetal04}, we have introduced a further subtype II$'_{abd}$ (whose definition is given in table~\ref{tab_BelDebever}), since it corresponds to an algebraic condition that is both simple (it gives type N in four dimensions) and geometrically relevant (see section~\ref{sec_cartan}). Additional subtypes can also be specified by more involved criteria or by combinations of those given above, but it seems of little interest to explore all possible subcases in full detail.

While various higher dimensional solutions (including, e.g., vacuum or charged supersymmetric black holes/rings) have been classified, and certain families of solutions defined by their algebraic type have been analyzed, several important issues deserves further study (see, e.g., \cite{Coley08} and references therein). It is hoped that the Bel-Debever criteria presented above will be helpful in future work. In what follows some applications are mentioned.

\section{Geometrical interpretation of WANDs: invariance under parallel transport}

\label{sec_cartan}

\subsection{Null vector}

In 1922 Cartan observed \cite{Cartan22} that in four dimensions the Weyl tensor defines at any spacetime point four privileged null directions, which have the property of being invariant under parallel transport around a certain (appropriately defined) family of infinitesimal parallelograms (see also \cite{RobRob72} for a comment on Cartan's paper). Here we further develop Cartan's description and extend it to higher dimensions, in order to provide a geometrical characterization of WANDs and (to some extent) of their multiplicity (with a natural, partial overlap with the previous section). In passing, we will also observe that the Petrov types of four dimensional vacuum spacetimes can be fully characterized this way (except for type D, for which properties of a second PND must also be considered).

Let $\bl$ be a null vector at a spacetime point $P$. If we are given another two (infinitesimal) vectors $\ep$ and $\del$ at $P$ (not parallel to each other), we can build an infinitesimal parallelogram with one vertex at $P$. Then, a basic result of Riemannian geometry tells us that if we parallely transport $\bl$ around such a parallelogram, when we are back at $P$ we will have, in general, a new vector $\bl'$ which differs from $\bl$ by $R^a_{\ bcd}l^b\epsilon^c\delta^d$ (see, e.g., \cite{Nakaharabook}). One can now ask how the Riemann tensor is restricted by requiring that: (i) the null direction defined by $\bl$ is preserved under parallel transport around the parallelogram, i.e. $R^a_{\ bcd}l^b\epsilon^c\delta^d=\lambda l^a$; (ii) the null vector $\bl$ is itself preserved, i.e. $R^a_{\ bcd}l^b\epsilon^c\delta^d=0$. These conditions can be imposed on various classes of parallelograms, resulting in different curvature restrictions. We will discuss this for five families of geometrically privileged parallelograms, defined by: 1) $\ep$ and $\del$ are, respectively, parallel and orthogonal to $\bl$ (i.e., $\ep=\gamma\bl$ and  $\del=\alpha\bl+\beta_i\m^{(i)}$ in a frame adapted to $\bl$); 2) $\ep$ is arbitrary but $\del$ is parallel to $\bl$; 
3) both $\ep$ and $\del$ are orthogonal to $\bl$; 4) both $\ep$ and $\del$ are arbitrary (generic parallelogram); 5)  $\ep$ is arbitrary but $\del$ is orthogonal to $\bl$. As it turns out, cases 4) and 5) give equivalent conditions, so that case 5) will be omitted hereafter. It is then straightforward to work out the necessary and sufficient conditions on the frame Riemann components for (i) or (ii) to happen. This is summarized in table~\ref{tab_parall}. Note that the conditions in column 4) are of particular geometrical meaning since they arise, e.g., in the study of spacetimes admitting a recurrent ($l_{a;b}=l_ap_b$) or covariantly constant ($l_{a;b}=0$) vector field (this, in turn, makes them relevant to discussions of holonomy, as we briefly exemplify below).

\begin{table}[t]
 \[
  \begin{array}{|l||l|l|l|l|} \hline
    & & & &  \\[-3mm]
    & \mbox{1) } \ep\parallel\bl , \ \del\perp\bl & \mbox{2) } \ep \mbox{ generic}, \del\parallel\bl & \mbox{3) } \ep\perp\bl , \ \del\perp\bl & \mbox{4) } \ep \mbox{ generic}, \del \mbox{ generic} \\
    & & & & \\[-3mm] \hline\hline
    & R_{0i0j} & R_{0i0j} & R_{0i0j} & R_{0i0j} \\
   \mbox{(i) } R^a_{\ bcd}l^b\epsilon^c\delta^d=\lambda l^a & \hfill  \mbox{I} & R_{010i} \hfill  \mbox{I$_a$} & R_{0ijk}   \hfill \mbox{II} & R_{010i}, R_{0ijk}   \hfill \mbox{II$_{abd}$} \\
   & & & & R_{0i1j} \ (R_{01ij}) \hfill \mbox{\scriptsize (X)} \\ 
      \hline
    & R_{010i} & & R_{010i} & \\
   \mbox{(ii) } R^a_{\ bcd}l^b\epsilon^c\delta^d=0 &  \hfill \mbox{I$_a$} & R_{0101} & R_{01ij}  \hfill \mbox{II$_d$} & R_{0101} \hfill \mbox{II$'_{abd}$} \\
    & & \hfill \mbox{\scriptsize (X)} & & R_{101i} \hfill \mbox{\scriptsize (X)} \\ \hline
  \end{array}
 \]
 \caption{\small~Frame components of the Riemann tensor that must vanish (at a spacetime point $P$) in order for the null direction defined by $\bl$ to be invariant under parallel transport around various families of infinitesimal parallelograms with one vertex at $P$ (the vanishing of such components is also a sufficient condition). The parallelograms are defined by the infinitesimal vectors $\ep$ and $\del$ at $P$ (specified in the upper row of the table), as explained in the main text. In row (ii) ($\lambda=0$) not only the null direction but also the vector $\bl$ is itself invariant and, in order to avoid repetition, it is understood that in each case the Riemann components that vanish in row (i) ($\lambda\neq 0$ case), same column, must still vanish. Riemann components in each box are given in order of decreasing boost weight, components with equal weights being on the same line (the component $R_{01ij}$, given in brackets, vanishes thanks to the cyclicity of the Riemann tensor). In each box it is also given the algebraic type of the corresponding Weyl tensor in the case of {\em vacuum} spacetimes ($R_{abcd}=C_{abcd}$) -- except in one special case, subcase of the type I$_a$, which does not correspond to any classified type. The same Weyl types also apply to {\em Einstein spaces} with $R\neq0$, except in the three cases marked by {\scriptsize (X)} (since $R_{0i1j}$ and $R_{0101}$ will differ from the corresponding Weyl components by terms proportional to $R$). (Note that in some cases the tracelessness of the Weyl tensor plays a crucial role in determining the type, so that not all of these types apply to the Riemann tensor of a generic, non-vacuum spacetime.)}
 \label{tab_parall} 
\end{table}
\begin{table}[h]
 \[
  \begin{array}{|l||l|l|l|l|} \hline
    & & & &  \\[-3mm]
    & \ep\parallel\bl , \ \del\perp\bl & \ep \mbox{ generic}, \del\parallel\bl & \ep\perp\bl , \ \del\perp\bl & \ep \mbox{ generic}, \del \mbox{ generic} \\
    & & & & \\[-3mm] \hline\hline
    & \Psi_0 & \Psi_0 & \Psi_0 & \Psi_0 \\
   C^a_{\ bcd}l^b\epsilon^c\delta^d=\lambda l^a & & \Psi_1 & \Psi_1 & \Psi_1 \\
   & \hfill \bl \mbox{ is a PND}& \hfill \mbox{II} & \hfill \mbox{II} & \Psi_2 \hfill \mbox{III} \\ \hline
    & \Psi_1 & & & \\
   C^a_{\ bcd}l^b\epsilon^c\delta^d=0 & & \Re(\Psi_2) & \Im(\Psi_2) & \\
    & \hfill \mbox{II} & & & \Psi_3 \hfill \mbox{N} \\ \hline
  \end{array}
 \]
 \caption{\small~The same as in table~\ref{tab_parall}, but in terms of the Weyl tensor, in the special case of $n=4$ spacetime dimensions. The equations in the first column determine which Weyl components (in Newman-Penrose notation) vanish, and viceversa. The vanishing of $\Psi_0$, by itself, does not tell anything about the Petrov type, but means that $\bl$ is a PND. Other conditions imply some algebraically special types, as specified in each box (when $\Re(\Psi_2)=0$ or $\Im(\Psi_2)=0$, but not both simultaneously, the type is still II). The equations in the first column are essentially equivalent to the standard Bel-Debever conditions. For instance, for type III one has $C^a_{\ bcd}l^b=\lambda_{cd}l^a$, which for type N reduces to $C^a_{\ bcd}l^b=0$ (since $\ep$ and $\del$ are arbitrary in those cases). In the particular case of vacuum spacetimes, the geometric interpretation discussed in the main text and in table~\ref{tab_parall} apply, and the Petrov types describe invariance properties of the corresponding PND(s) under parallel transport.}
 \label{tab_4dim}
\end{table}

Analogous conditions arise in terms of components of the Weyl tensor if we consider the equations $C^a_{\ bcd}l^b\epsilon^c\delta^d=\lambda l^a$ and $C^a_{\ bcd}l^b\epsilon^c\delta^d=0$, so that specific Weyl types are implied. For {\em vacuum} spacetimes one has $R_{abcd}=C_{abcd}$, and the interpretation of Weyl types in terms of properties of $\bl$ under parallel transport, as discussed above, applies (in particular, a generic WAND $\bl$ can be characterized as a null direction invariant under parallel transport around {\em all} infinitesimal parallelograms whose sides are parallel and orthogonal to $\bl$ itself). For a generic, non-vacuum spacetime this is no longer true since also the Ricci tensor will influence the parallel transport of $\bl$. If desired, it is easy to find under what restrictions on the matter content various conditions Weyl and Riemann components become equivalent. In particular, most of the types given in table~\ref{tab_parall} hold also for {\em Einstein spaces} (see table~\ref{tab_parall} for details). 
It turns out (cf.~the comments in the caption of table~\ref{tab_BelDebever}) that in four dimensions all Petrov types can be characterized this way (in vacuum). It it thus worth summarizing this in table~\ref{tab_4dim}.

While a comprehensive study of the relation between holonomy and Weyl types is well beyond the scope of this paper, let us just observe that the results of table~\ref{tab_parall} can be used to constraint the possible types allowed by certain holonomy groups. For instance, if the holonomy is (a subgroup of) Sim$(n-2)$, then there is a null direction invariant under parallel transport around any closed path. From row (i), column 4) of table~\ref{tab_parall} it then follows that in vacuum the Weyl type is II$_{abd}$ (or more special). For proper Einstein spaces\footnote{Higher dimensional Einstein spacetimes with holonomy Sim$(n-2)$ have recently been studied in \cite{GibPop08}.} one has $C_{0i1j}\propto R\delta_{ij}$ so that the type is only II$_{bd}$. For $n=4$ this means the Petrov types III and II, respectively, in agreement with the results of \cite{Schell61,GolKer61,Lewand92}. If the holonomy is instead (a subgroup of) E$(n-2)$, there is an invariant null vector and from row (ii), column 4)  we get the Weyl type II$'_{abd}$ in vacuum (proper Einstein spaces are now forbidden). In four dimensions this is just type N, which again agrees with \cite{Schell61,GolKer61,Lewand92}. 

\subsection{Timelike vector}

In the above discussion we have focused on invariance properties of a null vector under parallel transport. It is natural to investigate what happens if one instead considers vectors with non-zero length. We thus conclude the paper by briefly commenting on this case.

Let $\bt$ be a timelike vector. Due to norm conservation one has to discuss only the equation $R^a_{\ bcd}t^b\epsilon^c\delta^d=0$, since a non-null direction can not be otherwise preserved under parallel transport. We restrict here to the case when this hold for {\em any} infinitesimal parallelogram (other possibilities can be handled similarly). We then have
\be
 R^a_{\ bcd}t^b=0 .
 \label{timelike}
\ee 
We observe that this equation arises also in the study of spacetimes with a covariantly constant vector field $\bt$, i.e. when the holonomy is (a subgroup of) $SO(n-1)$, and it is thus of particular interest to clarify how it constraints the Weyl type.

Without loss of generality we can take $\bt$ to be normalized to $-1$. Now, {\em any} null vector must be of the form $\alpha(\bt\pm\bs)$, where $\bs\perp\bt$ is an appropriate spacelike vector (normalized to $+1$). In particular, we can always construct a null frame with $\bl=(-\bt+\bs)/\sqrt{2}$, $\bn=(\bt+\bs)/\sqrt{2}$, where $\bl$ is an arbitrary null vector (and $\bn$ its ``time-reflected''). Then eq.~(\ref{timelike}) is equivalent to $R_{abc0}=R_{abc1}$, which gives the following conditions on frame components 
\beqn
 R_{0i0j}=R_{1i1j}=R_{0i1j}=R_{1i0j}, \qquad R_{0i01}=0=R_{1i01}, \qquad R_{0ijk}=R_{1ijk}, \qquad R_{0101}=0 .
 \label{Riemann_timelike}
\eeqn

These equations constraint the Riemann type of any spacetime that admits a timelike vector satisfying~(\ref{timelike}) (one has also $R_{01ij}=0$ but this is not independent of~(\ref{Riemann_timelike})). More specifically, it is easy to see that the only possible Weyl types are G, I$_i$, D and O.\footnote{This is consistent with the result of \cite{PraPraOrt07} for static spacetimes. Note, indeed, that for any static spacetimes there exists a spacetime conformal to it that admits a covariantly constant timelike vector field (and thus obeys~(\ref{timelike})).}
 The Ricci tensor is also highly constrained and, in particular, proper Einstein spacetimes are not permitted (a contraction of~(\ref{timelike}) gives $R_{bd}t^b=0$).

Let us discuss in more detail the case of vacuum spacetimes, i.e., $R_{abcd}=C_{abcd}$. In general $\bl$ will not be a WAND, so that the type is G (algebraically general), but with the constraints~(\ref{Riemann_timelike}). In the case when $\bl$ is a WAND (i.e., $C_{0i0j}=0$), then by~(\ref{Riemann_timelike}) also $\bn$ is; the additional constraints now are $C_{0i1j}=0$ (or $\Phi_{ij}=0$ in the notation of \cite{PraPraOrt07}), $C_{0i01}=0=C_{1i01}$, $C_{0ijk}=C_{1ijk}$, and the type is I$_{ia}$ (meaning type I$_i$ and I$_a$, with respect to both WANDs, simultaneously). If, in addition, $\bl$ is a {\em mutiple} WAND (i.e., $C_{0ijk}=0$), then $\bn$ is multiple as well, and (since $\Phi_{ij}=0$) the type is D$_{abd}$. In four and five dimensions this actually means that the Weyl tensor vanishes and that the spacetime is flat (see \cite{Bel62,McIVan88} and the comments in the caption of table~\ref{tab_BelDebever}). However, this is not the case for $n>5$, since the Weyl part of $C_{ijkl}$ can still be non-zero.\footnote{For instance, non-flat vacuum spacetimes of type D$_{abd}$ can be easily constructed in $n\ge 6$ dimensions by taking a direct product of a Minkowski space with a curved but Ricci-flat Riemannian space (these are the \pp waves of type D mentioned in \cite{OrtPraPra09} -- cf.~the results of \cite{PraPraOrt07} for the properties of the Weyl tensor of direct product geometries). Less trivial examples are provided by the exceptional ``$\mu=0$'' Robinson-Trautman spacetimes \cite{PodOrt06} (see \cite{Ortaggio07,PraPraOrt07} for explicit solutions in any $n\ge7$ dimensions).} To summarize, only the algebraic types G, I$_{ia}$, D$_{abd}$ (and O) are compatible with~(\ref{timelike}) in vacuum.

Let us finally just mention that studying~(\ref{timelike}) when $\bt$ is spacelike does not seem to be of great interest in this context, in general. Indeed, given any Lorentzian metric (of any Riemann/Weyl type) one can add extra-dimensions by taking a direct product with a flat metric. Any spacelike vector with non-zero components only along the extra-dimensions will automatically satisfy~(\ref{timelike}) in the product geometry, without imposing any additional constraints on the Riemann/Weyl tensor (cf.~\cite{McIVan88} for $n=4$).

\section*{Acknowledgments}
I am grateful to Vojt\v ech Pravda for discussions and for reading the manuscript. Support from research plan No AV0Z10190503 and research grant KJB100190702 is acknowledged.

\addvspace{0.5cm}

%
%

\end{document}